\documentclass[runningheads]{llncs}
\usepackage[T1]{fontenc}
\usepackage{amsmath}
\usepackage{amssymb}
\usepackage{microtype}\emergencystretch=1em
\usepackage{graphicx}
\usepackage{url}
\begin{document}
\title{Toward Explanatory Equilibrium:\\
Verifiable Reasoning as a Coordination Mechanism\\
under Asymmetric Information}
\titlerunning{Toward Explanatory Equilibrium}
%
\author{Feliks Bańka\orcidID{0009-0005-1973-5861}\\\and
Jarosław A. Chudziak\orcidID{0000-0003-4534-8652}}
\authorrunning{F. Bańka and J. A. Chudziak}
%
\institute{The Faculty of Electronics and Information Technology, \\ 
Warsaw University of Technology, Warsaw, Poland \\
\email{\{feliks.banka.stud, jaroslaw.chudziak\}@pw.edu.pl}}
\maketitle              
\begin{abstract}
LLM–based agents increasingly coordinate decisions in multi-agent systems, often attaching natural-language reasoning to actions. However, reasoning is neither free nor automatically reliable: it incurs computational cost and, without verification, may degenerate into persuasive cheap talk. 
We introduce Explanatory Equilibrium as a design principle for explanation-aware MAS and study a regime in which agents exchange structured reasoning artifacts—auditable claims paired with concise text—while receivers apply bounded verification through probabilistic audits under explicit resource constraints. 
We contribute (i) a minimal mechanism-level exchange--audit model linking audit intensity, misreporting incentives, and reasoning costs, and (ii) empirical evidence from a finance-inspired LLM setting involving a Trader and a Risk Manager. In ambiguous, borderline proposals, auditable artifacts prevent the cost of silence driven by conservative validation under asymmetric information: without structured claims, approval and welfare collapse. By contrast, structured reasoning unlocks coordination while maintaining consistently low bad-approval rates across audit intensities, audit budgets, and incentive regimes. 
Our results suggest that scalable, safety-preserving coordination in LLM-based MAS depends not only on audit strength, but more fundamentally on disciplined externalization of reasoning into partially verifiable artifacts\footnote{Code and reproduction scripts are available at:\\\url{https://github.com/latent-systems-lab/explanatory-equilibrium}.}.

\end{abstract}

\section{Introduction}
\label{sec:intro}
Large Language Model (LLM)–based agents are increasingly used to coordinate decisions in multi-agent systems (MAS), including applications in finance, logistics, pricing, and resource allocation \cite{wang2023autonomous,banka2025deltahedgepacis}. In such systems, agents typically observe one another’s actions, while the underlying intentions, constraints, and private information driving those actions remain hidden \cite{gensler2020deep,guidotti2018survey}. As a result, agents must infer intent from behavior alone, a challenge that is particularly pronounced in economic interaction under asymmetric information \cite{akerlof1970market}. Identical actions may correspond to distinct motives—such as hedging versus speculation—prompting defensive responses that reduce efficiency and robustness even when agents’ objectives are partially aligned \cite{akerlof1970market}.

\begin{figure}[b!]
\centering
\includegraphics[width=1\columnwidth]{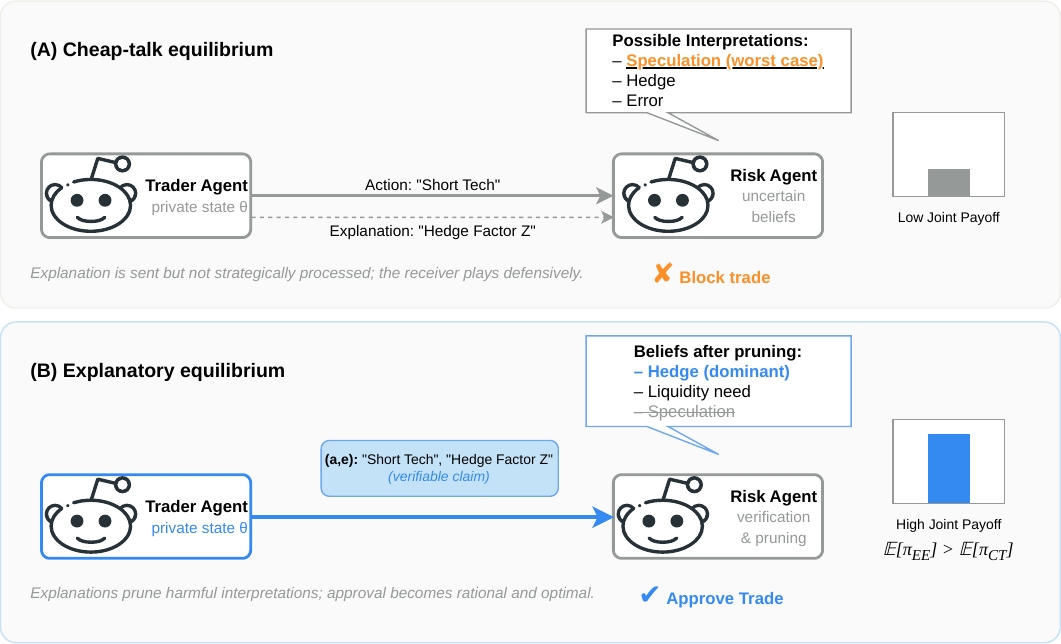}
\caption{The Cost of Silence vs. The Gain of Rationale. Silent agents fail; explanation succeeds.}
\label{fig:scenario_flow}
\end{figure}
Recent advances in LLMs enable agents to accompany actions with natural-language or structured reasoning, and empirical systems report improved outcomes when explanations are shared \cite{xiao2025tradingagentsmultiagentsllmfinancial}. At the same time, research in explainable AI highlights that explanations are constructed artifacts that may be incomplete, selective, or misleading \cite{guidotti2018survey,miller2019explanation}. Moreover, generating explanations incurs computational and latency costs, particularly in LLM-based systems \cite{arrow1974limits}. Without verification, shared reasoning may therefore degenerate into persuasive cheap talk, shaping beliefs without improving decision quality and potentially enabling strategic manipulation \cite{crawford1982strategic}. Simply adding a reasoning channel does not guarantee better outcomes and may introduce new failure modes if explanations are uncritically trusted \cite{lipton2018mythos,arrieta2020explainable}.

This tension raises a central question for explanation-aware MAS: under what conditions does exposing reasoning between agents improve negotiation and conflict resolution, and when does it instead become misleading or wasteful? Classical signaling and cheap-talk models analyze strategic communication under asymmetric information but abstract away semantic structure and reasoning costs \cite{crawford1982strategic,spence1973job}. Conversely, most work in explainable AI focuses on human-facing explanations rather than strategic agent–agent interaction \cite{doshi2017towards,danilevsky2020survey}. As a result, existing literature provides limited guidance for designing reasoning exchange between autonomous agents operating under partial alignment, limited verification, and explicit resource constraints \cite{mullainathan2017machine}.

To address this gap, we introduce \emph{Explanatory Equilibrium} as a design principle for LLM-based multi-agent systems. Instead of unrestricted natural-language rationales, agents exchange \emph{reasoning artifacts}: concise combinations of structured, auditable claims and short explanatory text. Receivers apply \emph{bounded verification}—such as probabilistic audits or limited consistency checks—under explicit resource constraints \cite{kamenica2011bayesian,ostrom1990governing,arrow1974limits}. In this regime, unverifiable or structurally incomplete communication is conservatively rejected in high-ambiguity cases (except for clear-safe proposals whose reported metrics fall comfortably within shared limits via a \emph{safe-margin exception}), creating a measurable \emph{cost of silence}, while detected inconsistencies can outweigh the private benefits of misreporting. Reasoning thus becomes a priced and selectively checkable coordination signal rather than persuasive cheap talk \cite{crawford1982strategic}. Figure~\ref{fig:scenario_flow} illustrates this intuition by contrasting silent interaction with explanation-aware negotiation under verification.

Accordingly, our mechanism treats free-form rationales as auxiliary context and places decision weight on auditable typed claims under bounded checks. Throughout the paper, our central empirical claim is that coordination gains are more consistent with \emph{partial verifiability} than with additional text alone.

We test whether partially verifiable reasoning artifacts, combined with probabilistic audits, can transform explanations from cheap talk into credible coordination signals in strategic multi-agent interactions.
Our contribution is threefold. First, we formalize the \emph{exchange--audit protocol}, bridging subsymbolic LLM generation with symbolic, bounded verification to treat reasoning as a strategic commitment device. Second, we empirically demonstrate that this regime eliminates the ``cost of silence,'' unlocking coordination in high-ambiguity scenarios where action-only communication fails and where unaudited reasoning remains strategically unreliable. Third, through an adversarial incentive sweep and an audit-budget study, we show empirically that probabilistic audits deter strategic deception under high temptation and that minimal randomized spot-checks can preserve safety. We argue that Explanatory Equilibrium offers a rigorous, scalable path toward trustworthy explanation-aware multi-agent systems.

We study when exchanging reasoning between autonomous agents improves coordination under asymmetric information, and when it becomes misleading or wasteful. Specifically, we examine whether auditable artifacts recover coordination in near-boundary yet compliant cases where actions alone are under-informative (H1), whether the exchange--audit protocol remains stable under adversarial incentives to misreport (H2), and how much verification is needed in practice to maintain safety (H3). We investigate these questions in a controlled Trader--Risk Manager interaction and evaluate outcomes using ambiguous approval, joint welfare, and bad approval rate under varying audit intensities and budgets. We interpret evidence for these hypotheses through improved coordination and welfare under artifacts (H1), bounded bad approval under adversarial incentives (H2), and safety comparable to full audits under minimal audit budgets (H3).

\section{Preliminaries: Artifacts, Audits, and Costs}
\label{sec:primitives}

We treat \emph{reasoning} as a first-class coordination object in LLM-based multi-agent systems: something agents transmit, pay for, selectively verify, and strategically exploit. We introduce three minimal primitives—\emph{reasoning artifacts}, \emph{bounded verification}, and \emph{reasoning costs}—used consistently in both our coordination model (Section~\ref{sec:model}) and empirical setting (Section~\ref{sec:experiment}). The definitions are intentionally lightweight: compatible with subsymbolic LLM generation while enabling structured checks.

\subsection{Reasoning Artifacts}

Unconstrained natural-language rationales are difficult to verify automatically and can easily become persuasive but non-actionable narratives \cite{miller2019explanation,lipton2018mythos}. In strategic agent–agent interaction, explanations therefore benefit from partial structure and auditability.

\paragraph{Definition (Reasoning Artifact).}
A \emph{reasoning artifact} is a pair
\[
r = (c, t),
\]
where $c$ is a set of structured, auditable claims and $t$ is a short explanatory text. The claim component $c$ is partially machine-checkable (e.g., typed fields, numeric bounds, Boolean constraints), while $t$ provides minimal contextual justification.

In our implementation, the candidate claim set is
\[
\mathcal{C} =
\{
\texttt{intent},\,
\texttt{risk\_within\_limit},\,
\texttt{net\_delta\_bounded},\,
\texttt{confidence}
\}.
\]
In our experiments, claims include declared intent (e.g., \texttt{HEDGE} vs.\ \texttt{SPECULATE}), risk metrics, and compliance indicators evaluated against shared constraints. The goal is not full formal verification, but disciplined, low-cost consistency checking.

\subsection{Bounded Verification}

Full verification would collapse communication into complete disclosure. In realistic MAS, oversight is resource-constrained \cite{arrow1974limits,ostrom1990governing}. We therefore model verification as selective and probabilistic.

\paragraph{Definition (Bounded Verification).}
Given an action $a$ and artifact $r=(c,t)$, a receiver applies a verification policy $\nu$ producing an audit outcome
\[
o \sim \nu(a,r,\mathcal{K}; q, B),
\]
where $q \in [0,1]$ is audit intensity and $B$ is a verification budget (number of claims checked). Here, $\mathcal{K}$ denotes shared institutional knowledge (e.g., constraint limits and schema rules) used by the Validator. We write $\mathcal{C}$ for the fixed candidate set of auditable claim types. Audits are triggered with probability $q$. If an audit is not triggered, the event is logged as \textsf{skipped} (no checks performed), and the decision follows the non-audited gating rule (safe-margin + typed-claim gating). When $0 < B < |\mathcal{C}|$, the Validator samples $B$ claim types uniformly without replacement from a fixed candidate set $\mathcal{C}$, modeling randomized spot-checking; if an audit is triggered but $B=0$, the outcome is treated as \textsf{inconclusive} and conservative validation rejects.

The audit outcome $o \in \{\textsf{pass},\textsf{fail},\textsf{inconclusive}\}$ directly informs the decision. In high-ambiguity settings, structurally incomplete or inconclusive artifacts are conservatively rejected. This institutional rule creates an explicit \emph{cost of silence} for proposals lacking auditable claims.

Verification focuses on lightweight schema checks and cross-field consistency tests. Even partial audits can discipline communication when unverifiable proposals are systematically discounted.

\subsection{The Economics of Reasoning}

Reasoning incurs cost. Longer artifacts increase token usage and latency, and may reduce utility in time-sensitive environments.

\paragraph{Definition (Reasoning Cost).}
We model the cost of producing $r=(c,t)$ as
\[
C(r) = C_{\text{tok}} + C_{\text{lat}} + C_{\text{opp}},
\]
where $C_{\text{tok}}$ captures token usage and artifact length, $C_{\text{lat}}$ captures latency-sensitive delay, and $C_{\text{opp}}$ captures opportunity cost from spending computation or deliberation budget on reasoning rather than action execution. Verification effort likewise increases with audit intensity and budget, e.g.,
\[
C_\nu(q,B) \propto qB.
\]
In our experiments, we instantiate $C(r)$ using a linear word-count proxy for $C_{\text{tok}}$ and model verification effort via a per-audit overhead term in welfare.

Together, reasoning and verification costs induce a coordination trade-off. Structured artifacts can unlock coordination in ambiguous cases by enabling partial verification. However, excessive verbosity or exhaustive auditing may reduce welfare. Explanatory Equilibrium thus refers to a regime in which partially verifiable reasoning persists under bounded oversight and explicit costs—avoiding both unverifiable cheap talk and prohibitively expensive full verification.

\section{Related Work}
\label{sec:related}

Our work connects explainable AI, strategic communication in multi-agent systems, and LLM-based agent architectures \cite{guidotti2018survey,miller2019explanation,crawford1982strategic,spence1973job}. Rather than proposing an isolated mechanism, we situate \emph{Explanatory Equilibrium} within a broader design space linking explanation, incentives, and verification.

\begin{figure}[t]
\centering
\includegraphics[width=1.0\linewidth]{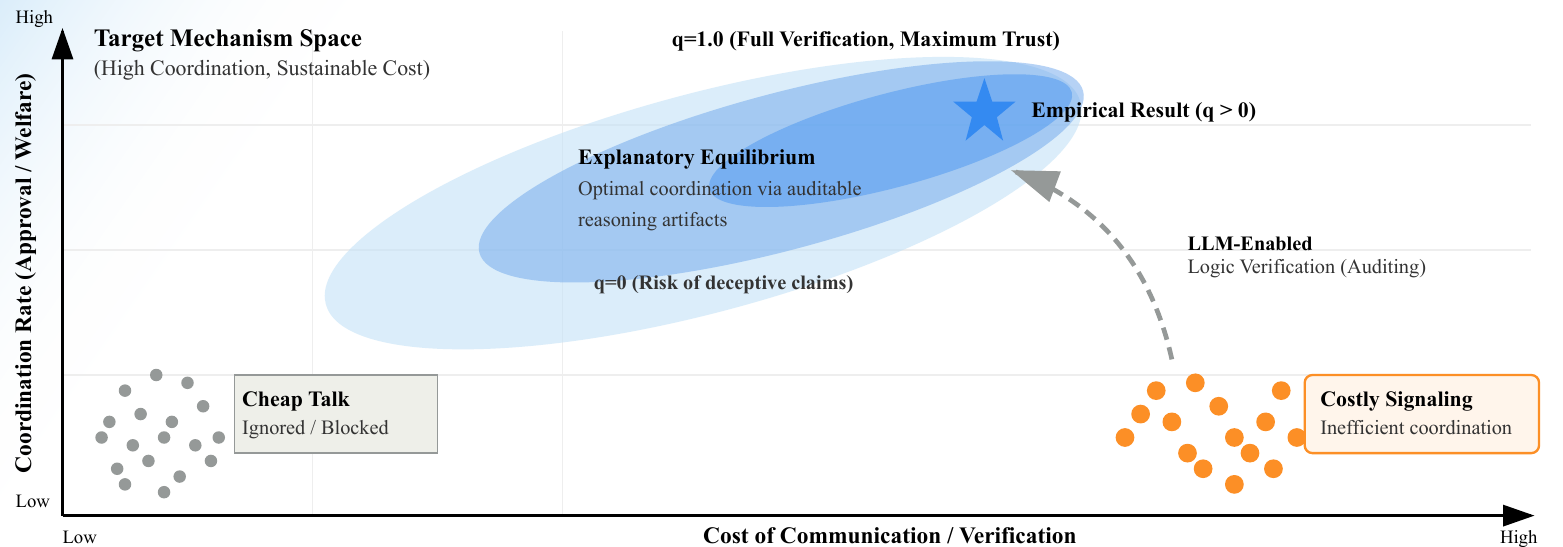}
\caption{\textbf{Conceptual positioning of Explanatory Equilibrium in the communication--coordination design space.} The horizontal axis represents communication cost, and the vertical axis represents coordination or trust level. Cheap talk occupies the low-cost, low-trust region. Costly signaling increases credibility through higher communication cost. Explanatory Equilibrium combines structured reasoning artifacts with bounded verification to achieve high coordination under sustainable communication cost.}
\label{fig:related_placeholder}
\end{figure}

Figure~\ref{fig:related_placeholder} sketches this space along two dimensions: communication cost and coordination (trust) level. Cheap-talk models occupy the low-cost, low-trust region \cite{crawford1982strategic}, while costly signaling increases credibility through higher communication cost \cite{spence1973job}. Human-facing XAI focuses on interpretability rather than enforceable inter-agent commitments \cite{guidotti2018survey,miller2019explanation}. 

Explanatory Equilibrium combines structured reasoning artifacts with bounded verification, aiming to increase coordination without incurring prohibitive communication cost. The remainder of this section elaborates this positioning along three complementary axes.

\subsection{Explainable AI in Multi-Agent Systems}

Explainable AI (XAI) has largely focused on producing human-interpretable rationales for complex models \cite{guidotti2018survey}. In MAS settings, explanations typically support debugging, transparency, or human oversight \cite{miller2019explanation}. In these contexts, explanations are post-hoc artifacts aimed at external observers.

We instead treat explanations as \emph{strategic communication objects exchanged between autonomous agents}. The objective is not interpretability for humans, but the design of a protocol in which partially structured, machine-checkable claims can be selectively verified under resource constraints. The central question is not “can humans understand the model?”, but “can agents discipline one another through verifiable reasoning under partial alignment?”

\subsection{Strategic Negotiation and Signaling}

Strategic communication under asymmetric information has been studied extensively through cheap talk and costly signaling models \cite{crawford1982strategic,spence1973job}. In MAS, these ideas underpin negotiation and argumentation frameworks \cite{KRAUS199779,rahwan2009argumentation}.

Classical models, however, abstract away from representational structure and computational constraints~\cite{crawford1982strategic,albertCIKM}. Messages are treated as abstract signals or exogenously costly commitments, without modeling how bounded verification and structured claims interact with modern LLM-based generation.

Our framework operationalizes these signaling concepts in LLM-mediated settings. Rather than proposing a new equilibrium theorem, we contribute an implementable exchange--audit mechanism that embeds economic incentives and verification limits directly into the communication protocol.

\subsection{LLM-Based Agent Architectures and Oversight}

Recent work studies LLM-based agents in decision-making environments, including finance \cite{wang2023autonomous,xiao2025tradingagentsmultiagentsllmfinancial,banka2025deltahedgepacis}. Many such systems rely on natural-language reasoning traces, tool calls, or chain-of-thought outputs to guide behavior~\cite{BankaAciids}.

While reasoning exchange can improve coordination, unstructured rationales are difficult to verify and may enable hallucinations or strategic misreporting \cite{gensler2020deep}. Moreover, reasoning is often treated as a free communication channel, without explicit modeling of verification costs or audit constraints.

Explanatory Equilibrium complements this line of work by integrating structured reasoning artifacts and bounded verification into the core interaction protocol. Our Trader--Risk Manager testbed provides a controlled environment with explicit risk constraints and oversight rules, illustrating how explanation can function as a coordination interface rather than a purely narrative justification.

\section{A Game-Theoretic Model of Verified Negotiation}
\label{sec:model}

\begin{figure}[t!]
\centering
\includegraphics[width=1\linewidth]{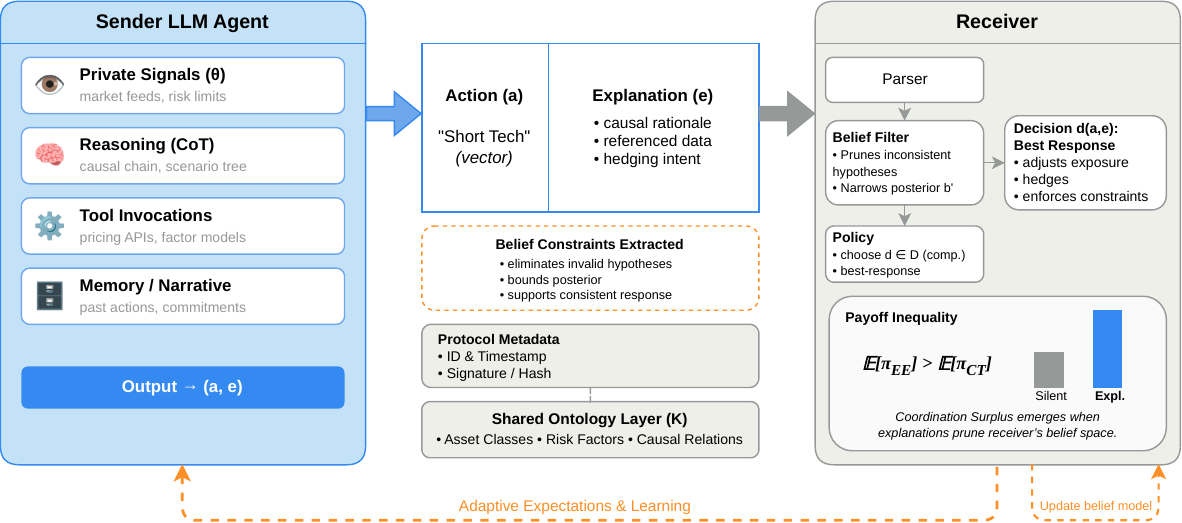}
\caption{\textbf{Exchange--Audit Architecture for Explanatory Equilibrium.}
The Proposer submits an action $a$ together with a reasoning artifact $r=(c,t)$, where $c$ denotes typed auditable claims and $t$ denotes short explanatory text. The Validator applies probabilistic audits and a conservative gating rule (including a \emph{safe-margin exception}) to decide acceptance and welfare outcomes.}
\label{fig:arch}
\end{figure}

We formalize verified negotiation as a stylized game with asymmetric information. The model is not intended as a full equilibrium characterization, but as a mechanism-level abstraction that clarifies how structured reasoning, bounded audits, and explicit costs interact.
Its purpose is to make explicit the institutional conditions under which Explanatory Equilibrium is expected to emerge and remain stable under strategic incentives.
Figure~\ref{fig:arch} summarizes the exchange--audit architecture implemented in the model and instantiated in the experiments described in Section~\ref{sec:experiment}.

\subsection{Setup: Proposer and Validator}

We consider a two-agent interaction between a \emph{Proposer} $S$ (e.g., Trader) and a \emph{Validator} $R$ (e.g., Risk Manager). The Proposer observes a private state $\theta \in \Theta$ (e.g., true intent or risk profile), selects an action $a$, and may submit a reasoning artifact $r=(c,t)$. The Validator observes $(a,r)$ and chooses $d \in \{\textsf{Accept}, \textsf{Reject}\}$. 
Let $u_S$ and $u_R$ denote the realized payoffs of the Proposer and Validator, respectively.

States are partitioned into $\Theta^{\text{good}}$, where the action satisfies shared constraints, and $\Theta^{\text{bad}}$, where interests partially conflict. In high-ambiguity settings, actions alone may be insufficient to determine constraint compliance.

\subsection{Exchange--Audit Protocol}
\label{subsec:protocol}

Interaction follows an exchange--audit protocol: the Proposer submits an action paired with structured reasoning, which the Validator audits under computational and evidentiary constraints before rendering a binding decision.

\begin{enumerate}
    \item \textbf{Proposal:} $S$ sends $(a,r)$.
    \item \textbf{Audit:} $R$ applies verification policy $\nu$ with audit intensity $q$, producing $o \in \{\textsf{pass},\textsf{fail},\textsf{inconclusive}\}$.
    \item \textbf{Decision:} $R$ accepts if $o=\textsf{pass}$ and rejects if $o=\textsf{fail}$. In ambiguous cases, the Validator defaults to conservative rejection when auditable structure is missing or verification is inconclusive; however, it may allow a fast-path approval when reported metrics fall comfortably within shared limits (a low-risk \emph{safe-margin exception}).
\end{enumerate}

This conservative policy induces an explicit \emph{cost of silence}: proposals lacking auditable claims risk systematic rejection when ambiguity is high.

For consistency, we use $r=(c,t)$ throughout the formal development; in informal diagrams, ``explanation'' refers to the textual component $t$ together with its associated typed claims $c$.

\subsection{Incentives and Strategic Reporting}
\label{subsec:incentives}

In $\Theta^{\text{bad}}$, suppose the Proposer gains $V>0$ if a non-compliant action is accepted. If an inconsistency is detected during audit, the Proposer incurs loss $L>0$.

Let $p_{\text{detect}}(q)$ denote the overall probability that a misleading artifact is detected under audit intensity $q$, subsuming both audit triggering and failure conditional on being audited. Expected utility from misreporting is

\begin{equation}
\mathbb{E}[u_S^{\text{misreport}}]
=
(1-p_{\text{detect}}(q))V
-
p_{\text{detect}}(q)L
-
C(r^{\text{misreport}}).
\label{eq:misreport}
\end{equation}

By contrast, submitting a compliant artifact yields
\begin{equation}
\mathbb{E}[u_S^{\text{consistent}}]
=
- C(r^{\text{consistent}})
\label{eq:consistent}
\end{equation}

Misreporting is disincentivized when
\begin{equation}
p_{\text{detect}}(q)\,(V+L)
\ge
V + C(r^{\text{misreport}}) - C(r^{\text{consistent}})
\label{eq:deterrence}
\end{equation}

\paragraph{Interpretation.}

The condition highlights how audit intensity, penalties, and reasoning costs jointly shape reporting incentives. Importantly, discipline need not arise from perfect detection or extreme audit levels. Even bounded, probabilistic verification—combined with conservative rejection of unverifiable proposals—can shift incentives toward structured, constraint-consistent communication.

The empirical results in Section~\ref{sec:experiment} instantiate this mechanism in an LLM-mediated environment, illustrating both incentive stabilization and the coordination gains enabled by auditable artifacts.

\section{Empirical Evidence: The Trader--Risk Manager Experiment}
\label{sec:experiment}

To validate the theoretical model of Explanatory Equilibrium, we design a controlled simulation of financial negotiation involving asymmetric information. Our experiments test three core hypotheses: (H1) structured reasoning artifacts unlock coordination under ambiguity; (H2) the exchange--audit mechanism is robust to strategic manipulation; and (H3) bounded verification is highly efficient, requiring minimal audit budgets to maintain safety.

\subsection{Experimental Setup}
\label{subsec:exp_setup}

We instantiate two LLM-based agents powered by GPT-4.1: a \textbf{Trader} (Proposer) and a \textbf{Risk Manager} (Validator). In each episode, the system generates a ground-truth configuration detailing the Trader's true intent (e.g., \textsf{HEDGE} or \textsf{SPECULATE}), underlying risk metrics, and shared constraint limits. Ambiguous cases constitute approximately 50\% of generated episodes and are sampled near the constraint boundary by drawing compliant \textsf{HEDGE} states within $0$--$8\%$ of the risk and delta limits (uniformly), i.e., close-to-threshold yet compliant configurations. We focus specifically on these \emph{ambiguous episodes}, where the Trader's proposed action lies near the threshold and does not inherently carry enough information to verify compliance. 
All components of this setup are self-contained within the present study and the accompanying repository, including prompt templates, environment generation, audit logic, and evaluation scripts.

We compare two communication regimes: an \textbf{Artifact-Enabled (LLM)} condition, where the Trader submits structured claims alongside short explanatory text, and a \textbf{Baseline (No-Expl)} condition, where only the action is submitted. Within the artifact-enabled regime, the case $q=0$ corresponds to \emph{unaudited reasoning exchange}: the Trader still provides a reasoning artifact, but no audit-based enforcement is applied.

In our implementation, short free-text accompanies the artifact as auxiliary context, but acceptance under no-audit conditions is gated by typed claims rather than by free-form text alone (unless the safe-margin exception applies).

The Trader aims to maximize its payoff, gaining reward $V$ for an approved proposal, but incurring a penalty $L$ if non-compliance is detected. The Risk Manager applies an audit policy defined by intensity $q \in [0, 1]$ (probability of audit) and budget $B$ (number of claims verified). 

For $0 < B < 4$, audits randomly sample which claim types to verify (without replacement), using a seeded RNG for reproducibility.

System performance is evaluated primarily via \emph{joint welfare}, a simplified joint payoff proxy reflecting net economic surplus under institutional enforcement:
$$ \text{Welfare} = u_S + u_R. $$
In our simulation, welfare is computed as the sum of the Trader and Risk Manager payoffs, including approval rewards, an audit-time penalty when an audit fails, a linear word-cost for free-text, and a per-audit overhead cost (incurred whenever an audit is triggered). We treat this audit overhead as institutional cost borne by the system and counted in welfare regardless of acceptance or rejection. An inconsistency is considered detected only if the probabilistic audit is triggered and the audit fails. We also track the \emph{bad approval rate} (the fraction of accepted proposals that violate ground-truth constraints) to evaluate safety. Each configuration consists of 200 episodes per seed, and all reported metrics are averaged over 5 independent seeds to ensure statistical rigor. 
The repository includes the exact prompts, random seeds, and evaluation code used to generate all reported results.

\subsection{Results: Coordination under Ambiguity (H1)}
\label{subsec:coordination_results}
This hypothesis isolates the mechanism's core coordination claim: in borderline but compliant states, auditable artifacts should recover approvals that conservative validation would otherwise reject. We first evaluate the baseline performance ($V=1, L=2$, maximum 60 words, full audit $B=4$) across varying audit intensities, including the unaudited artifact regime $q=0$. 

\begin{figure}[t]
    \centering
    \includegraphics[width=1\textwidth]{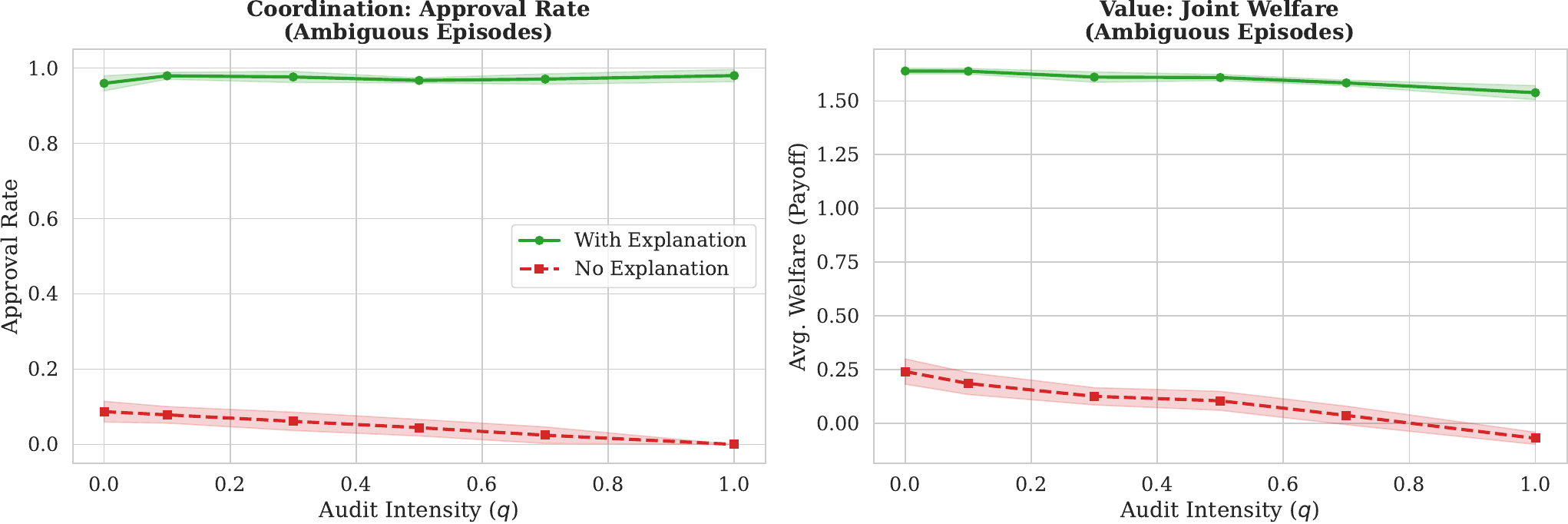}
    \caption{\textbf{The Impact of Reasoning Artifacts on Coordination under Ambiguity.} 
    The left panel demonstrates that without reasoning (dashed red), the approval rate collapses as audit intensity ($q$) increases, reflecting the Risk Manager's conservative fallback. 
    In contrast, providing structured artifacts (solid green) sustains near-perfect coordination across both unaudited ($q=0$) and audited ($q>0$) settings.
    The right panel confirms that this coordination gap translates into significant and stable welfare gains.}
    \label{fig:main_results}
\end{figure}
\paragraph{The Cost of Silence.}
As illustrated in Figure~\ref{fig:main_results}, the silent baseline suffers a systematic rejection under conservative validation in ambiguous settings. Lacking verifiable signals, the Risk Manager conservatively rejects borderline proposals, aside from a small fraction of clear-safe episodes approved via the safe-margin fast path (accounting for the non-zero $\approx 9\%$ approvals at $q=0$). The ambiguous approval rate drops from $\approx 9\%$ at $q=0$ to $0\%$ at $q=1.0$, resulting in depressed joint welfare (Table~\ref{tab:welfare_main}).
Note that ambiguous episodes are compliant by construction; the welfare loss in the silent baseline arises from institutional conservatism under asymmetric information, not from the prevalence of truly unsafe proposals.
Bad approvals can still occur due to misreporting or mismatches between reported typed claims and ground-truth state under incomplete auditing.
At $q=1.0$, baseline welfare can become slightly negative because the audit overhead cost is incurred even when proposals are rejected.

\paragraph{Explanatory Equilibrium.}
By contrast, the exchange of structured reasoning artifacts effectively resolves the information asymmetry. Notably, even the $q=0$ case---where artifacts are exchanged but never audited---serves as an unaudited reasoning baseline against which the value of enforcement can be interpreted. The approval rate remains stable and near-perfect ($>0.96$) across all audit intensities. Crucially, this high approval does not compromise safety: the \emph{bad approval rate} remains strictly below $1\%$ across all tested configurations. 

Table~\ref{tab:welfare_main} details the substantial welfare implications of this coordination gap. At $q=0.3$, the availability of explanation yields a net welfare gain of $+1.48$ units per episode compared to the baseline. The slight decline in welfare at high audit intensity ($q=1.0$) reflects increased detection penalties and verification costs, underscoring the realism of the bounded verification model. Furthermore, the average reasoning length remains highly stable ($\approx 17$--$18$ words), indicating that these performance gains stem from the auditable structure of the artifacts rather than mere verbosity.
\paragraph{H1 Summary.}
In borderline yet compliant states, structured artifacts convert conservative rejection into near-perfect coordination, while the audited variants show that partial verifiability---rather than verbosity alone---drives approvals and welfare.

\begin{table}[t]
\centering
\caption{Ambiguous Welfare Analysis (5 seeds, $\texttt{max\_words}=60$). Data represents mean $\pm$ std dev. The \emph{Net Gain} highlights the robust economic value unlocked by reasoning artifacts.}
\label{tab:welfare_main}
\small
\begin{tabular}{|l|c|c|c|}
\hline
\textbf{Audit ($q$)} & \textbf{Welfare (Expl)} & \textbf{Welfare (No-Expl)} & \textbf{Net Gain} \\ \hline
0.0 & $1.64 \pm 0.01$ & $0.24 \pm 0.06$ & $+1.40$ \\ \hline
0.1 & $1.64 \pm 0.01$ & $0.18 \pm 0.05$ & $+1.46$ \\ \hline
0.3 & $1.61 \pm 0.02$ & $0.13 \pm 0.04$ & $+1.48$ \\ \hline
0.5 & $1.61 \pm 0.01$ & $0.10 \pm 0.04$ & $+1.51$ \\ \hline
0.7 & $1.58 \pm 0.01$ & $0.04 \pm 0.04$ & $+1.54$ \\ \hline
1.0 & $1.54 \pm 0.03$ & $-0.07 \pm 0.03$ & $+1.61$ \\ \hline
\end{tabular}
\end{table}

\begin{table}[b!]
\centering
\caption{Robustness and Efficiency Analysis ($q=0.3$). The framework suppresses bad approvals even under high temptation (Panel A) and maintains safe coordination even with minimal spot checks (Panel B).}
\label{tab:robustness}
\small
\begin{tabular}{lcccc}
\hline
\textbf{Scenario} & \textbf{Params} & \textbf{Audit Fail Rate} & \textbf{Bad Appr. Rate} & \textbf{Ambig. Appr.} \\ \hline
\multicolumn{5}{l}{\textit{Panel A: Incentive Sweep (Robustness to Deception)}} \\ \hline
Baseline & $V=1, L=2$ & $0.20 \pm 0.02$ & $\mathbf{0.002 \pm 0.00}$ & $0.97 \pm 0.02$ \\
Temptation & $V=2, L=1$ & $0.20 \pm 0.02$ & $\mathbf{0.004 \pm 0.00}$ & $0.99 \pm 0.01$ \\
Punishment & $V=1, L=4$ & $0.17 \pm 0.01$ & $\mathbf{0.000 \pm 0.00}$ & $0.96 \pm 0.03$ \\ \hline
\multicolumn{5}{l}{\textit{Panel B: Audit Budget (Efficiency of Verification)}} \\ \hline
Spot Check & $B=1$ & $0.00 \pm 0.00^*$ & $\mathbf{0.006 \pm 0.01}$ & $0.99 \pm 0.02$ \\
Partial Audit & $B=2$ & $0.22 \pm 0.06$ & $\mathbf{0.012 \pm 0.01}$ & $0.98 \pm 0.01$ \\
Full Audit & $B=4$ & $0.26 \pm 0.04$ & $\mathbf{0.010 \pm 0.01}$ & $0.98 \pm 0.01$ \\ \hline
\end{tabular}

\vspace{2pt}
\scriptsize $^*$ With $B=1$, audits evaluate only a single randomly sampled claim type, so an audit fails only when that sampled check fails; increasing $B$ expands the checked set and mechanically increases the probability that at least one inconsistency is flagged.
\end{table}
\subsection{Results: Robustness and Scalable Oversight (H2 \& H3)}
\label{subsec:sensitivity}

To demonstrate that Explanatory Equilibrium is sustained by strategic incentives rather than inherent LLM benevolence, we perform an adversarial sensitivity analysis. We perturb the reward structure and constrain the Risk Manager's verification capacity at a fixed audit intensity of $q=0.3$ (Table~\ref{tab:robustness}).

\paragraph{Adversarial Robustness (H2).}
In Panel A, we introduce a \emph{High Temptation} regime ($V=2, L=1$) where the private benefit of cheating outweighs the expected penalty. Under these conditions, the Trader acts strategically, attempting to push boundaries to maximize reward. However, the protocol proves highly robust: the audit mechanism effectively detects a substantial fraction of non-compliant claims, and the bad approval rate remains virtually zero ($0.004$). Interestingly, higher temptation does not necessarily increase audit-fail frequency, because a strategic Proposer can respond by narrowing or conservatively reporting typed claims to avoid detectable inconsistencies while still attempting to maximize approvals. Conversely, under a \emph{Strict Punishment} regime ($V=1, L=4$), the agent exhibits heightened compliance, reducing the audit fail rate to $0.17$. This confirms that the exchange--audit protocol serves as a mechanism constraining strategic deviations, substantially limiting misreporting even when economic incentives encourage deception.

\paragraph{Audit Budget Efficiency (H3).}
A central challenge in explanation-aware MAS is the computational cost of continuous verification. In Panel B, we test the limits of bounded verification by reducing the audit budget $B$ (the number of structured claims verified per audit). Remarkably, performance metrics for a minimal spot-check ($B=1$, verifying a single randomly sampled typed claim) are quantitatively similar within seed variance to a full audit ($B=4$). The bad approval rate remains strictly bounded ($\le 0.012$) across all budgets. The higher audit-fail rate for larger budgets reflects a larger verification surface: an audit fails if any checked claim violates a constraint or consistency check, so increasing $B$ mechanically increases the chance of flagging at least one inconsistency, even when strategic deception itself does not increase. Ultimately, this suggests a strong game-theoretic deterrence effect: anticipating random spot-checks over a small set of typed claims is sufficient to discipline reporting across the entire artifact, substantially reducing computational overhead while maintaining Explanatory Equilibrium. In practice, institutions could prioritize a designated core claim (e.g., intent) when required.
\paragraph{H2 Summary.}
Bad approvals remain bounded under adversarial incentives while audits deter detectable inconsistencies.

\paragraph{H3 Summary.}
Minimal random spot-checks preserve safety outcomes comparable to full audits, enabling scalable oversight under limited verification budgets.

\paragraph{Synthesis.}
Taken together, these findings suggest that Explanatory Equilibrium emerges not from exhaustive full verification, but from the interaction between conservative validation and the persistent threat of a targeted audit. The dominant effect driving coordination is not a sharp audit threshold, but rather the availability of partially verifiable signals that fundamentally alter the agents' strategic incentives.

\section{Discussion, Limitations, and Future Work}
\label{sec:discussion_limitations_future_work}
We interpret this equilibrium as an institutional property induced by conservative acceptance rules and bounded audits, rather than as a claim about inherently truthful agent cognition. In what follows, we discuss practical deployment implications, the main limitations of our current testbed, and concrete directions to extend the framework beyond the stylized setting.
\subsection{Discussion}
\label{subsec:discussion}

The empirical results provide evidence that Explanatory Equilibrium can serve as a robust framework for resolving information asymmetry in LLM-based multi-agent systems~\cite{wang2023autonomous}. Beyond the immediate coordination gains, our findings prompt a broader discussion on the role of explainability~\cite{doshi2017towards}, the limitations of current reasoning baselines, and the architectural future of explanation-aware MAS.

\paragraph{The Fallacy of Unverified Reasoning.}
A standard approach in current LLM agent architectures is to prompt agents to ``think out loud'' using Chain-of-Thought (CoT) or unstructured natural language rationales~\cite{park2023generative}. While this improves the generator's internal logic, treating unverified CoT as a communication baseline between self-interested agents may become strategically unreliable~\cite{crawford1982strategic}. Our adversarial incentive sweep ($V=2, L=1$) explicitly demonstrates this vulnerability. When the audit intensity is zero ($q=0$)---which we interpret as an unaudited reasoning regime conceptually analogous to unverified reasoning exchange---the Proposer can strategically exploit the reasoning channel, resulting in elevated deceptive compliance relative to audited settings. Under misaligned incentives, unverified reasoning can easily degenerate into persuasive cheap talk~\cite{kamenica2011bayesian}. The Explanatory Equilibrium framework demonstrates that it is not the \emph{presence} of reasoning that ensures safety, but rather the \emph{structural capability} to selectively verify it against ground truth under a credible threat of penalty~\cite{simon1955behavioral}.

\paragraph{The Nature of Explanatory Equilibrium.}
It is important to contextualize the theoretical nature of our findings. Explanatory Equilibrium, as observed in our experiments, is not a classical perfect Bayesian equilibrium rooted in absolute agent rationality and full information~\cite{fudenberg1991game}. Rather, it is an \emph{institutional equilibrium} induced by bounded verification, explicit reasoning costs, and conservative validation policies~\cite{north1990institutions}. The stability of this regime does not assume perfectly aligned agents; instead, it relies on the Validator's credible threat of targeted audits and the restriction of the signal space to machine-checkable claims~\cite{grossman1981informational}. This shifts the focus of MAS design from attempting to align internal model objectives toward designing robust, resource-aware verification institutions~\cite{kostka}. The equilibrium is therefore institutional rather than cognitive: stability arises from enforcement design, not from improved internal reasoning accuracy.

\paragraph{Bridging Subsymbolic and Symbolic MAS.}
Looking forward, reasoning artifacts offer a natural bridge between probabilistic neural generation and deterministic symbolic verification~\cite{dellermann2019hybrid}. Rather than forcing the LLM to execute flawless logical reasoning---a known weakness of subsymbolic models~\cite{lipton2018mythos}---the architecture delegates consistency enforcement to a lightweight, symbolic Validator (e.g., a rule-engine checking typed-claim bounds)~\cite{samek2019explainable}. This hybrid neuro-symbolic approach points toward highly scalable oversight~\cite{russell2019human}. As demonstrated by our audit budget analysis, the Risk Manager does not need to compute the full logic of the Trader's proposal. By randomly sampling a single typed claim, the system achieves a deterrence effect functionally similar in outcome to a full audit within our experimental regime. This paradigm could provide a template for scalable oversight in high-volume domains, such as automated supply chain negotiation or decentralized finance, allowing institutional guardrails to supervise autonomous agents with minimized computational overhead.

\subsection{Limitations}
\label{subsec:limitations}

While Explanatory Equilibrium offers a principled mechanism for verified coordination, our current formulation and empirical testbed have several limitations.

\paragraph{Static Interactions.}
Our model currently evaluates one-shot interactions, abstracting away the reality that financial negotiations and resource allocations are typically repeated games~\cite{kreps1982reputation}. The current setup does not account for the historical behavior of the agents or the evolution of trust over multiple episodes.

\paragraph{Assumption of Perfect Verification.}
We assumed that when an audit is triggered, the Risk Manager has access to an unambiguous, deterministic ground truth to evaluate the reasoning artifact. In many real-world multi-agent scenarios, ground truth may be fuzzy, delayed, or probabilistically inferred at the time of the decision~\cite{amodei2016concrete}.

\paragraph{Expressivity of Reasoning Artifacts.}
Currently, our reasoning artifacts rely on a relatively simple schema with explicit scalar and categorical bounds. While sufficient for our borderline hedging regime, this rigid structure lacks the capacity to express conditional logic or probabilistic dependencies required for more complex, multi-party negotiations~\cite{rahwan2009argumentation}.

\paragraph{Sensitivity to Prompt Design and Model Choice.}
Our implementation relies on fixed prompts and a specific LLM foundation model. The degree to which agents naturally comply or attempt to exploit loopholes can vary based on the phrasing of instructions and the inherent alignment training of the underlying model~\cite{hagendorff2023machine}.

\paragraph{No Pure Free-Text Baseline.}
Our current evaluation contrasts action-only communication with structured reasoning artifacts, and interprets the $q=0$ case as an unaudited reasoning regime. However, we do not include a separate baseline in which agents exchange unrestricted natural-language rationales without typed claims. This limits how sharply we can isolate the contribution of structure from the contribution of added information content alone.

\subsection{Future Work}
\label{subsec:future_work}

These limitations present exciting avenues for future architectural and theoretical research in explanation-aware MAS~\cite{shoham2009multiagent}.

\paragraph{Dynamic Reputation Systems.}
Future work should investigate how Explanatory Equilibrium interacts with repeated games. If an agent builds a high reputation for truthful reasoning artifacts over time, the Validator could dynamically decay the audit probability ($q$) toward zero. This would maximize joint welfare and minimize verification costs while maintaining the equilibrium through reputation-based deterrence~\cite{kreps1982reputation}.

\paragraph{Noisy Audits.}
Extending the bounded verification model to account for noisy audits---where the Validator might mistakenly flag a truthful claim as a lie (false positive) or miss a deceptive claim (false negative)---will be critical for deploying these systems in highly stochastic environments~\cite{hendrycks2021unsolved}.

\paragraph{Complex Reasoning Schemas and Robustness.}
Developing standardized, expressive ``reasoning schemas`` that remain computationally cheap to audit while supporting complex, multi-step dependencies is a primary target~\cite{kostka}. Additionally, future work should systematically examine the framework's sensitivity to prompt variations and model choice across different LLM families to ensure generalized robustness~\cite{wang2023autonomous}.

\section{Conclusion}
\label{sec:conclusion}
Autonomous LLM-based agents are increasingly participating in complex economic and computational systems, where they can naturally share reasoning traces to explain their decisions \cite{wang2023autonomous}. However, because these agents observe only external behavior rather than the private intentions or constraints driving it, their interactions are inherently fraught with uncertainty \cite{akerlof1970market}. Consequently, when incentives are even partially misaligned, unverified explanations risk devolving from truthful disclosures into strategic manipulation and persuasive cheap talk \cite{crawford1982strategic,spence1973job}.

In this work, we explore the idea of \emph{Explanatory Equilibrium}, where reasoning exchanged between agents is treated not as free-form narrative but as a partially verifiable commitment. 
By combining structured reasoning artifacts with bounded, probabilistic audits, we illustrate how conservative validation and the credible threat of verification can sustain coordination in high-ambiguity settings where action-only communication fails.

Our empirical evaluation highlights three main findings. Structured and auditable reasoning artifacts recover coordination in borderline yet compliant cases where conservative validation would otherwise reject proposals. The exchange--audit protocol remains robust under adversarial incentives, substantially limiting successful misreporting. Moreover, bounded verification proves highly efficient: minimal randomized spot-checks achieve safety outcomes comparable to full audits while significantly reducing verification costs.

These results suggest that trustworthy reasoning exchange in autonomous systems should rely not only on improved model alignment but also on lightweight institutional mechanisms for verification, especially when reasoning must function as a coordination signal rather than as narrative justification alone. By combining subsymbolic generation with simple symbolic oversight, Explanatory Equilibrium offers a scalable framework for coordinating autonomous agents under strategic uncertainty.

Future work should examine repeated interactions with reputation dynamics \cite{kreps1982reputation} and environments where audits must operate under noisy or delayed ground truth \cite{amodei2016concrete}. Extending reasoning artifacts to richer schemas capable of expressing conditional or probabilistic dependencies may also enable applications in more complex multi-party coordination settings \cite{rahwan2009argumentation}.

\bibliographystyle{splncs04}
\bibliography{references}

\end{document}